\documentclass[conference,a4paper]{IEEEtran}
\IEEEoverridecommandlockouts
\usepackage{cite}
\usepackage{amsmath,amssymb,amsfonts}
\usepackage{algorithmic}
\usepackage{graphicx}
\usepackage{textcomp}
\usepackage{xcolor}
\def\BibTeX{{\rm B\kern-.05em{\sc i\kern-.025em b}\kern-.08em
    T\kern-.1667em\lower.7ex\hbox{E}\kern-.125emX}}
\begin{document}

\title{
A Comparative Analysis of Ising Formulations for Neuromorphic Maximum-Likelihood Channel Decoding}

\author{\IEEEauthorblockN{George N. Katsaros, Morgan Sabine and Konstantinos Nikitopoulos}
\IEEEauthorblockA{6G Innovation Centre \\
University of Surrey, UK}\\
% City, Country \\
% email address or ORCID
}

\maketitle

\begin{abstract}
Neuromorphic computing has so far been driven predominantly by machine-learning workloads, yet its underlying properties also make it particularly well suited to combinatorial optimization problems expressed in Ising or QUBO form. 
While neuromorphic Ising solvers have been demonstrated, how a given problem should be formulated to best suit neuromorphic dynamics has received far less attention.
Maximum-likelihood (ML) channel decoding can be expressed as an Ising/QUBO problem, and two distinct formulations already exist in the quantum-annealing literature: a squared-penalty formulation that uses few spins but produces dense intra-check couplings, and a chain-product formulation that improves locality at the cost of additional auxiliary spins. Both place the ML codeword at the ground state under sufficient constraint enforcement, but they have not been compared under the constraints that neuromorphic hardware imposes. This work provides the first systematic side-by-side
comparison of QUBO/Ising formulations of ML decoding for linear codes. We show that the two formulations impose fundamentally different tradeoffs in neuron count, synaptic density, locality, and convergence behavior. The preferred formulation is inseparable from the choice of solver, and the two must be considered jointly. Finally, we show that ground-state correctness alone is an insufficient design criterion, and that signal processing tasks should ideally be co-formulated with their neuromorphic hardware models if neuromorphic computing is to extend into the receiver pipeline.
\end{abstract}

% \begin{IEEEkeywords}
% Neuromorphic, Channel Coding, Ising
% \end{IEEEkeywords}

% ============================================================
\section{Introduction}
% ============================================================

Forward-error correction (FEC), particularly channel decoding, remains one of the most computationally demanding operations in the receiver chain \cite{mansour_640-mbs_2006}. 
While maximum-likelihood (ML) decoding is optimal, its direct realization entails an exhaustive codeword search and therefore becomes impractical as the code length increases.
Consequently, practical receivers rely on iterative message-passing algorithms, such as belief propagation (BP). However, these decoders on conventional digital hardware are typically memory-bound and power-hungry \cite{guan_check-belief_2023}. Their repeated message exchanges demand substantial memory access and interconnect bandwidth, challenging energy-constrained receivers.

To overcome such computational and architectural bottlenecks, a recent growing body of research has explored the use of physical solvers for combinatorial optimization. Taking inspiration from how  physical systems evolve toward states of minimal energy, these hardware substrates leverage their intrinsic dynamics to minimize an objective function. Most of these solvers are natively described using quadratic unconstrained binary optimization (QUBO) or equivalent Ising models. In a QUBO formulation, the objective is expressed as a quadratic function of binary variables, whereas equivalently the Ising form maps the same problem onto spin variables interacting via local fields and pairwise couplings. The broad class of physical solvers encompasses quantum and probabilistic annealers \cite{Johnson2011QuantumAW}, oscillator-based Ising machines \cite{twang_osci_ising_2019}, as well as neuromorphic systems \cite{neuromimo,spawcneuromimo}. Among these, neuromorphic architectures appear particularly promising.
Unlike many emerging physical computing substrates, neuromorphic architectures do not rely on breakthroughs in materials science for practical realization, as they can be implemented using established CMOS technologies.
Thus, they provide a readily available and highly scalable hardware foundation that has shown significant promise for ultra-low-power and massively parallel computation \cite{furber_large-scale_2016}.

Despite rapid hardware advances in neuromorphic computing platforms, these developments have not yet been matched by an equally mature set of algorithmic frameworks, especially for problems beyond the scope of machine-learning \cite{aimone_reviewnoncogn_2022}. This gap suggests that the next major challenge is not simply to build larger neuromorphic platforms or more efficient neuron circuits, but to determine how specific algorithms can be reformulated to exploit neuromorphic principles in a meaningful way. 
This is particularly true for channel decoding, where multiple Ising/QUBO mappings exist in the quantum-annealing literature \cite{kasi_towards_2020,ide_maximum_nodate} but have never been compared under the constraints neuromorphic hardware imposes.

As we discuss later in detail, although ML decoding of a binary linear code can be expressed as the minimization of a channel-dependent objective under parity-check constraints, making this problem compatible with an Ising/QUBO Hamiltonian requires these constraints to be incorporated into the objective through a chosen penalty construction. 
Therefore, multiple Ising/QUBO mappings can encode the same underlying ML decoding problem. They differ, however, not only in how reliably the ML codeword is represented as the ground state (i.e., the lowest-energy state) of the resulting Hamiltonian, but also in the hardware and solver requirements they impose, and these differences become visible only when the mappings are compared directly.

In this work, we present the first systematic side-by-side comparison of QUBO/Ising formulations of ML decoding, originally introduced for quantum-based solvers, when the target implementation platform is neuromorphic.
We treat the mathematical formulation as a key design variable and ask: given the ML decoding problem, how does the chosen mapping reshape the physical energy landscape, and what are the implications for the solver attempting to navigate it?
We examine and compare two representative approaches (A) a squared-penalty construction that is highly spin-economical but induces dense inter-neuron couplings \cite{kasi_towards_2020}, and (B) a chain-product construction that is connectivity-economical but expands the search space by introducing auxiliary spin chains~\cite{ide_maximum_nodate}. 

Specifically, we show that ground-state correctness alone is an
insufficient criterion for assessing the practical viability of an Ising/QUBO formulation on neuromorphic hardware. A formulation must
also be evaluated in terms of the tradeoffs it creates between neuron
count, synaptic density, locality, landscape navigability, and
compatibility with the solver dynamics. 
In particular, our results show that minimizing the number of neurons is
not necessarily the most hardware-efficient choice, since it can shift the
cost to denser coupling graphs and increased routing complexity.
Conversely, formulations that reduce coupling density and improve locality
may do so by introducing auxiliary neurons, thereby increasing the size of
the search space.
Moreover, while stronger parity-constraint enforcement enables the ML codeword to become the lowest-energy state, fewer nearby states lead the solver toward it, shrinking the ML basin of attraction significantly.
Lastly, we highlight that for the channel coding, the neural coupling topology can be fixed per code-rate and be computed offline while only the self-biases update per received block. This eliminates the runtime preprocessing overhead that a conventional host would otherwise incur. 
Together, these observations shift the focus from whether ML decoding can be mapped to an Ising Hamiltonian to how such mappings should be designed with respect to the target hardware. In doing so, they provide practical guidelines for neuromorphic hardware designers and a concrete starting
point for the co-design of neuromorphic FEC accelerators.

\section{System Model}\label{sec:system}
% Rewrite

Consider a binary linear code of length $n$ that is defined by a parity-check matrix $\mathbf{H}\in \mathbb{F}_2^{m \times n}$ such that a valid codeword $\mathbf{x}\in \mathbb{F}_2^{n \times 1}$ satisfies $\mathbf{H}\mathbf{x}\equiv\mathbf{0}\pmod{2}$. Each bit is BPSK modulated to a transmitted symbol through the transformation $1-2x_i\in\{1,-1\}$. After transmission over an AWGN channel with noise variance $\sigma^2$, the received symbol is
$y_i=(1-2x_i)+n_i,$ where $n_i\sim\mathcal{N}(0,\sigma^2).$
In maximum-likelihood (ML) decoding, the codeword that maximizes the probability $P(\mathbf{y|x})$ is sought. Taking the logarithm of the factorized Gaussian likelihood and exploiting the $x_i^2=x_i$ property reduces the squared Euclidean distance to a linear form, yielding the constrained optimization
\begin{equation}\label{eq:ml}
  \hat{\mathbf{x}}
  = \arg\min_{\mathbf{x}}
    \sum_{i=1}^{n} L_i\, x_i
  \quad\text{s.t.}\quad
  \mathbf{H}\mathbf{x}\equiv\mathbf{0}\!\!\pmod{2},
\end{equation}
where $L_i=2y_i/{\sigma^2}$ is the log-likelihood ratio (LLR) for the $i$th bit. Crucially, the channel contribution is entirely linear in $\mathbf{x}$. Consequently, the circuit topology remains fixed across codeblocks, and, as we discuss in Section \ref{sec:programming}, only the diagonal entries need to be updated for each received block.

Each row $c$ of $\mathbf{H}$ defines a parity check over the variable set $\mathcal{N}(c)=\{j:[\mathbf{H}]_{cj}=1\}$, with check degree $d_c=|\mathcal{N}(c)|$, such that a valid codeword satisfies $\sum_{j\in\mathcal{N}(c)} x_j \equiv 0 \pmod{2}$, thus imposing an explicit constraint on the ML decoding problem. Therefore, to obtain an unconstrained formulation amenable to Ising hardware, the parity constraint must be incorporated into the Ising Hamiltonian as a penalty. The next section presents two such penalty constructions.

\section{Ising Formulations for Channel Decoding}\label{sec:formulations}

Both formulations start from the ML objective~\eqref{eq:ml} and produce the standard Ising Hamiltonian

\begin{equation}\label{eq:ising}
  \mathcal{H}(\mathbf{s})
  = -\!\sum_{i<j} J_{ij}\,s_i s_j
    - \sum_{i} h_i\,s_i,
\end{equation}

where the $i$th spin $s_i\in\{-1,+1\}$ has a pairwise coupling $J_{ij}$ to the $j$th spin, and a local bias $h_i$. However, the two formulations differ fundamentally in how they treat the parity-check constraint, leading to distinct spin counts, coupling densities, and energy landscape characteristics.

\subsection{Squared-Penalty Formulation}\label{sec:A}

Because check $c$ requires $\sum_{j\in\mathcal{N}(c)} x_j$ to be even, there exists a non-negative integer $k_c\in\{0,1,\ldots,\lfloor d_c/2 \rfloor\}$ such that the constraint is equivalent to $\sum_{j\in\mathcal{N}(c)} x_j = 2k_c$. To penalize any deviation from this equality, the following penalty term is introduced:
\begin{equation}\label{eq:penalty_A}
  P_c
  = \Bigl(\sum_{j\in\mathcal{N}(c)} x_j - 2k_c\Bigr)^{\!2}.
\end{equation}
Since QUBO form requires all variables to be binary, $k_c$ is encoded as $B_c=\lceil\log_2(\lfloor d_c/2\rfloor+1)\rceil$ auxiliary bits $\mathbf{a}_c=(a_{c,0},a_{c,1},\cdots,a_{c,B_c-1})\in \mathbb{F}_2^{1 \times B_c}$ such that $k_c=\sum_{l=0}^{B_c-1}2^la_{c,l}$. Expanding~\eqref{eq:penalty_A} produces only linear and pairwise terms (since $x_j^2=x_j$ and $a_{c,l}^2=a_{c,l}$). Concatenating all $n$ code bits and all auxiliary bits into a single vector $\mathbf{z}=[\mathbf{x},\mathbf{a}_1,\mathbf{a}_2,\ldots ,\mathbf{a}_m]^T\in \mathbb{F}_2^{N \times 1}$ where $N=n+\sum_{c=1}^mB_c$, the unconstrained objective Hamiltonian becomes

\begin{equation}\label{eq:qubo}
  \mathcal{H}_A
  = \sum_{i=1}^{n}L_i\,x_i
    + \lambda\sum_{c=1}^{m}P_c
  = \mathbf{z}^{T}\!\mathbf{Q}\,\mathbf{z},
\end{equation}
where $\mathbf{Q}$ is a real symmetric $N\!\times\!N$ matrix and
$\lambda>0$ controls the penalty strength.
The QUBO can then be mapped to the Ising form~\eqref{eq:ising} via the
substitution $x_i=(1-s_i)/2$, which yields couplings
$J_{ij}=-Q_{ij}/4$ and local biases
$h_i = -q_i/2 - \tfrac{1}{4}\sum_{j\neq i}Q_{ij}$,
where $q_i=Q_{ii}$.

\subsection{Chain-Product Formulation}\label{sec:B}

Approach B constructs the Ising Hamiltonian directly in the spin domain. Using the mapping $s_j=(-1)^{x_j}$, the parity-check condition for check $c$ is represented by the spin product

\begin{equation}\label{eq:product}
  \prod_{j\in\mathcal{N}(c)} s_j
  =
  (-1)^{\sum_{j\in\mathcal{N}(c)} x_j},
\end{equation}
which evaluates to $1$ when the number of participating bits is even, and $-1$ otherwise. Thus, a natural penalty is
\begin{equation}\label{eq:multibody_penalty}
  P_c
  =
  \frac{\lambda}{2}
  \left(
  1-\prod_{j\in\mathcal{N}(c)}s_j
  \right),
\end{equation}
which equals zero for a satisfied check and $\lambda$ for a violated check. However, this formulation results in a $d_c$-body interaction, whereas a standard Ising Hamiltonian permits only linear and pairwise terms. To reduce the multi-body product, product-accumulator spins $p_{c,l}\in\{-1,1\}$ can be introduced, where
\begin{equation}\label{eq:chain_def}
    p_{c,1} \triangleq s_{c,1}, \qquad
    p_{c,l} = p_{c,l-1} \cdot s_{c,l}, \quad
    l=2,3, \ldots,d_c,
\end{equation}
recursively, such that the terminal product spin $p_{c,d_c}$ is the chain product $p_{c,d_c}=\prod_{j\in\mathcal{N}(c)}s_j$.
Therefore, the parity contribution to the Ising Hamiltonian can be written as the linear term
\begin{equation}\label{eq:parity_B}
  \mathcal{H}_P
  =
  \frac{\lambda_{P}}{2}
  \left(
  m-\sum_{c=1}^{m}p_{c,d_c}
  \right).
\end{equation}

While this term rewards terminal product spins equal to $+1$, the Ising machine does not automatically enforce the relations in \eqref{eq:chain_def}. Without additional constraints, the product spins may decouple from the code-bit spins, allowing the terminal variables $p_{c,d_c}$ to minimize \eqref{eq:parity_B} without correctly representing the parity of the corresponding check.

To enforce each relation in \eqref{eq:chain_def}, a penalty is introduced using an additional auxiliary spin $a_{c,l}\in\{-1,1\}$. A four-spin penalty function $\mathcal{H}_C(p_{c,l},p_{c,l-1},s_{c,l},a_{c,l})$ is constructed such that when the relation $p_{c,l}=p_{c,l-1} \cdot s_{c,l}$ is satisfied, there exists an assignment of $a_{c,l}$ yielding $\mathcal{H}_C=0$, and a penalty $\mathcal{H}_C>0$ when the relation is violated. One such construction, given in~\cite{ide_maximum_nodate}, is
\begin{equation}\label{eq:Hcl}
\begin{aligned}
  \mathcal{H}_{C} ={}& \tfrac{1}{2}\bigl(
        p_{c,l}\,p_{c,l-1} + p_{c,l-1}\,s_{c,l} + p_{c,l}\,s_{c,l}
      \bigr) \\
      &+ \bigl(a_{c,l} + \tfrac{1}{2}\bigr)
      \bigl(2a_{c,l} - p_{c,l} - p_{c,l-1} - s_{c,l}\bigr).
\end{aligned}
\end{equation}
The complete Hamiltonian is therefore
\begin{equation}\label{eq:full_B}
\begin{aligned}
    \mathcal{H}_B
    &=
    -\frac{1}{2}\sum_{i=1}^nL_is_i+\frac{\lambda_P}{2}\left(m-\sum_{c=1}^mp_{c,d_c}\right)\\
    &\qquad+\lambda_C\sum_{c=1}^m \sum_{l=2}^{d_c} \mathcal{H}_C\left(p_{c,l},p_{c,l-1},s_{c,l},a_{c,l}\right),
\end{aligned}
\end{equation}
where $\lambda_P$ and $\lambda_C$ control the enforcement strength of the parity-check and chain link, respectively. Since all terms in \eqref{eq:full_B} are linear or pairwise, the formulation is already in standard Ising form.
Finally, it's worth mentioning here that other higher-order (HUBO) constructions exist that retain the $d_c$-body parity term but they fall outside the standard 2-local Ising form assumed by current neuromorphic and annealer hardware, and are therefore beyond the scope of this work.

\section{Energy Landscape Analysis}\label{sec:landscape}

\begin{figure}[t]
    \centering
    \includegraphics[]{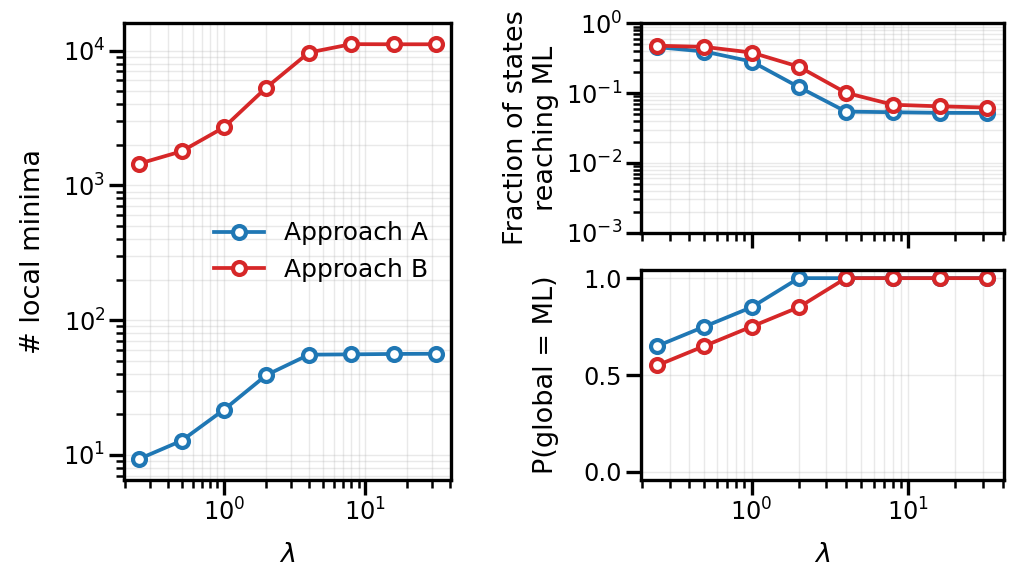}
\caption{Exact landscape statistics for the $(2,4)$-regular code
($M=3$, $n=6$), comparing Approach~A ($N_A=12$ spins, blue) and
Approach~B ($N_B=24$ spins, red) as a function of the
constraint-enforcement parameter~$\lambda$, averaged over 100 channel
realisations at $E_b/N_0=2$\,dB. Left: mean number of single-flip
local minima. Top right: fraction of initial states from which greedy
single-flip descent reaches the ML codeword.
Bottom right: probability that the global energy minimum coincides with
the ML codeword.}
    \label{fig:num_local_minima}
    % \vspace{-8pt}
\end{figure}

In this section, we analyze the energy landscapes induced by the two Ising formulations, assessing their exactness and the convergence challenges that practical solvers encounter due to the constraint-enforcement parameter $\lambda$.
Fig. \ref{fig:num_local_minima} shows the exact energy landscape statistics for a small $(2,4)$-regular code with parity-check matrix $\mathbf{H}\in\mathbb{F}_2^{3\times6}$, chosen to keep exhaustive enumeration feasible. This yields $N_A=12$ and $N_B=24$ total spins for approaches A and B, respectively, corresponding to  $2^{N_A}=4096$ and $2^{N_B}\approx1.68\times10^7$ possible configurations. For both approaches, we enumerate all spin configurations for each value of $\lambda$ and compute their Ising energy. Results are averaged over 100 channel realizations at an SNR of $2~\text{dB}$. The ML reference is obtained independently via an exhaustive search over all valid codewords.

Fig.~\ref{fig:num_local_minima} (bottom-right) shows the probability that the global minimum matches the ML codeword as a function of $\lambda$. For weak constraint enforcement, parity-violating configurations can have lower energy than the ML codeword because a favorable channel outweighs the penalty. As $\lambda$ increases, these invalid configurations are increasingly penalized, and the probability that the global minimum and ML are equal approaches one. However, stronger constraint enforcement also fragments the energy landscape. Fig.~\ref{fig:num_local_minima} (left) shows that the number of local minima grows with $\lambda$ for both formulations, with approach B exhibiting substantially more local minima than approach A across all $\lambda$ values, attributed to its larger spin space. In both cases, the majority of additional minima ($>70\%$) do not correspond to valid configurations. The penalty terms introduce couplings that reshape the local energy surface, so a parity-violating configuration, while not globally favorable, may still be stable with respect to every single-spin flip. 

The practical consequence is visible in the basin of attraction of the ML solution. For each possible initial configuration, we run greedy single-spin-flip descent until a local minimum is reached. Fig. \ref{fig:num_local_minima} (top-right) shows that while increasing $\lambda$ improves global exactness, it sharply reduces the fraction of initial states from which local descent converges to the ML solution. In other words, although the correct solution becomes the deepest point in the energy landscape, its basin of attraction occupies a progressively smaller fraction of the state space. Fine-tuning $\lambda$ should therefore be jointly optimized with the choice of parity-check matrix, Ising formulation, and solver strategy. In addition,  as we discuss in Section \ref{sec:solvers}, this observation motivates solver initialization from channel-consistent configurations, such as the hard-decision estimate, together with stochastic search mechanisms that can escape shallow local minima.
From a neuromorphic standpoint, the depth of these spurious minima sets a lower bound on the perturbation amplitude a spiking solver must inject, recasting $\lambda$ as a joint formulation–noise design parameter.

\section{Solver Strategies}\label{sec:solvers}

In this section, we evaluate the performance of deterministic and stochastic Ising-based solvers against the channel hard-decision baseline and a standard sum-product belief propagation (flooding schedule, 50 iterations) using a $(3,6)$-regular LDPC code with parity-check matrix $\mathbf{H}\in\mathbb{F}_2^{48\times 96}$. The hard-decision decoder uses only the signs of the channel log-likelihoods, whereas the Ising decoders minimize an energy function that includes the same channel term, alongside a parity-check penalty. The two Ising-based approaches are presented in Section \ref{sec:formulations}, and the decoded word is obtained solely from the code-spin variables.

\begin{figure}
    \centering
    \includegraphics[]{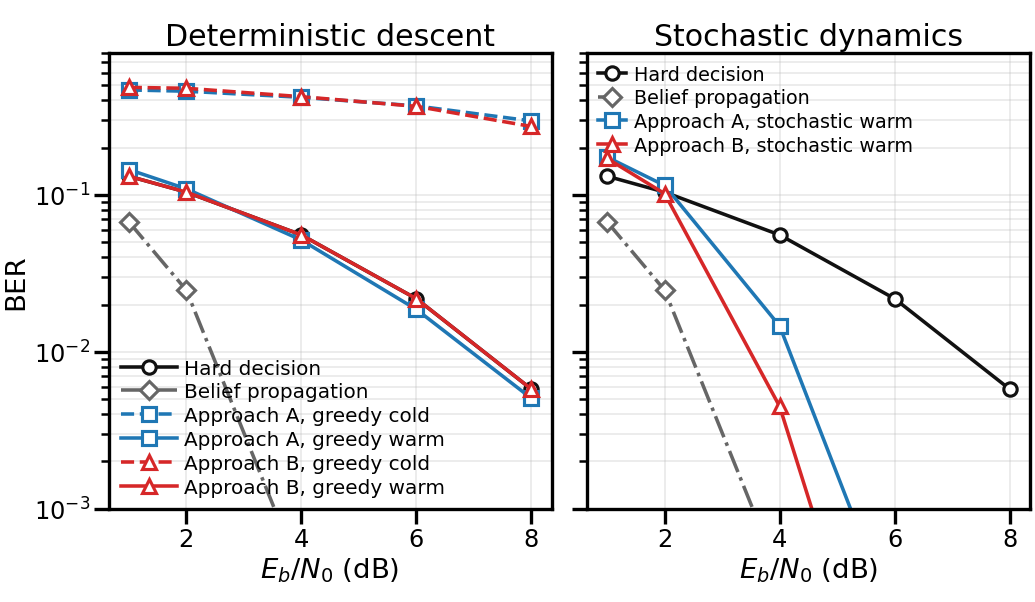}
     \caption{BER vs.\ $E_b/N_0$ for the $(3,6)$-regular LDPC code
    ($48\times 96$). Left: greedy single-spin-flip descent with cold start (random initialisation) and hard-decision warm start for
    both Approach~A ($\lambda=3$) and Approach~B
    ($\lambda_P=8$, $\lambda_C=1.5$).   
    Right: simulated annealing with 
    $3000$ annealing steps, selecting the lowest-energy outcome. The hard-decision and BP
    baselines are common to both panels.}
    \label{fig:berresults}
    % \vspace{-15pt}
\end{figure}

% ------------------------------------------------------------
\subsection{Greedy Descent}%
\label{sec:warmstart}
% ------------------------------------------------------------
The deterministic solver uses single-spin greedy descent, similar in operation to the neuromorphic greedy-descent QUBO solver utilized for MU-MIMO detection\cite{spawcneuromimo,neuromimo}. It is evaluated under both cold- and warm-start initialization. In the cold-start case, the solver begins from a random spin configuration. In the warm-start case, the code spins are initialized to the channel hard decision, while the auxiliary spins are initialized according to the corresponding codeword. This has the advantage of starting the search in a physically meaningful region of the energy landscape. 

Specifically, in approach A, the auxiliary register for check $c$ encodes the integer $k_c$ appearing in the squared penalty. Setting $S_c$ as the hard-decision check sum, we set $k_c=S_c/2$ when $S_c$ is even. When $S_c$ is odd, the two nearest integers $\lfloor S_c/2\rfloor$ and $\lceil S_c/2 \rceil$ are both equally valid, providing a natural multi-start extension. 
For approach B, the product accumulator spins are initialized by evaluating the running parity product $p_{c,l}=p_{c,l-1}s_l$. Additionally, the auxiliary spin $a_{c,l}$ is initialized such that $H_C=0$. These auxiliary initializations are local per-check computations that scale as $O(d_c)$ and require no global linear solve or matrix inversion. This is a practical advantage of Ising-based LDPC decoding since the channel LLRs provide a natural
initial code-word estimate, and the auxiliary variables can be filled in locally. However, a consistent warm start is not a guarantee of greedy improvement. As we discuss below, in Approach~B, the same chain consistency that removes artificial auxiliary energy can also make the hard-decision state a local minimum for single-spin descent.

Fig.~\ref{fig:berresults} (left) shows the resulting BER curves.
Cold-started greedy descent performs poorly for both formulations. 
Random initial spin configurations almost always converge to an invalid
local minimum. Hard-decision warm starting
significantly improves the picture, but the gains relative to the
hard-decision remain marginal. 
For Approach A, a hard-decision bit flip can reduce the squared parity residuals of its incident checks, so warm-started greedy descent can improve slightly on the hard decision baseline
when the parity gain exceeds the channel cost.  
The situation is worse for the greedy descent when utilizing Approach~B.
The reason is structural. At initialisation, every chain and link
constraint is exactly satisfied, so the only penalty contribution comes
from the parity term $\lambda_P$ at checks where the hard
decision violates parity. To correct such a check, at least one of its
code spins must be flipped. However, each code spin $s_i$ participates
in $d_v$ checks, and flipping it violates the chain
constraint $p_m = p_{m-1}\,s_i$ in every one of those checks. The
resulting energy increase is of order $\lambda_C\,d_v$,
because the accumulator spins $p_m$ now disagree with the flipped code
spin. The solver therefore rejects the flip, and the hard-decision state
remains a local minimum. Reducing $\lambda_C$ to lower this
barrier is not viable since the parity term reads the final accumulator
spin $p_{c,d_c}$, which only represents the true check parity when
the chain is intact. Weakening the chain enforcement in turn corrupts
the very mechanism that should guide the solver toward valid codewords.
Convergence in Approach~B therefore requires accepting an initially
uphill code-spin flip followed by a cascade of accumulator repairs, a trajectory that greedy descent cannot initiate.

\subsection{Stochastic Solvers}\label{sec:stochastic}
To show the effect of stochastic search in overcoming the abovementioned energy barriers, we run simulated annealing (SA) from the same hard-decision-consistent warm state. We use the classical SA implementation from D-Wave's
\texttt{dwave-neal} package that can act as a lower-bound of typical neuromorphic Ising solvers.  The solver receives the Ising
Hamiltonian $(\mathbf{J},\mathbf{h})$ for each formulation and is initialized
from the same hard-decision warm start used in the greedy experiments. The experiments follow the default geometric inverse-temperature schedule.
Each decoding instance runs $20$ independent annealing trajectories of
$3000$ steps each, with the lowest-energy outcome selected as the
decoded word. 
The penalty weights were selected via a coarse grid search at
$4$~dB SNR minimizing BER, yielding $\lambda = 3$ for
approach~A and $(\lambda_P, \lambda_C) = (8, 1.5)$ for approach~B.
Fig.~\ref{fig:berresults} (right) shows that a stochastic search substantially
improves performance for both formulations, closing much of the gap to BP for approach~A and recovering meaningful coding gain for approach~B, which was entirely trapped at the hard-decision under greedy descent.

These results motivate the employment and design of neuromorphic Ising solvers, which
can natively exhibit the stochastic dynamics that SA emulates algorithmically. 
Spiking neural networks can naturally support
Metropolis-like exploration through random perturbations
of the membrane potential that cause neurons to fire or remain silent
independently of their deterministic input, producing spontaneous state
transitions equivalent to uphill spin flips. Gradually reducing the
noise amplitude implements a cooling schedule analogous to SA,
allowing the network to transition from broad exploration to
convergence. 
Additional mechanisms explored in the literature involve perturbing the neuron refractory periods in which after firing, a neuron is temporarily unable
to fire again, and thus forcing exploration of new regions of the
state space \cite{jonke_solvingcspsnn}.
It is important to mention here that the present results establish that the formulations' energy
landscapes are navigable by the class of stochastic dynamics
neuromorphic solvers natively provide. While a direct neuromorphic implementation is left for future work, recent work has shown that neuromorphic solvers can achieve performance comparable to SA on sparse Ising graphs~\cite{Henke2024ComparingQA}, which is precisely the regime that channel-coding Ising formulations inhabit owing to the sparse structure of the parity-check matrix (see Section~\ref{sec:sparsity} and Table~\ref{tab:comparison}).

\section{Sparsity and Complexity Tradeoffs}\label{sec:sparsity}

\begin{figure}[h]
    \centering
    \includegraphics[height=5cm]{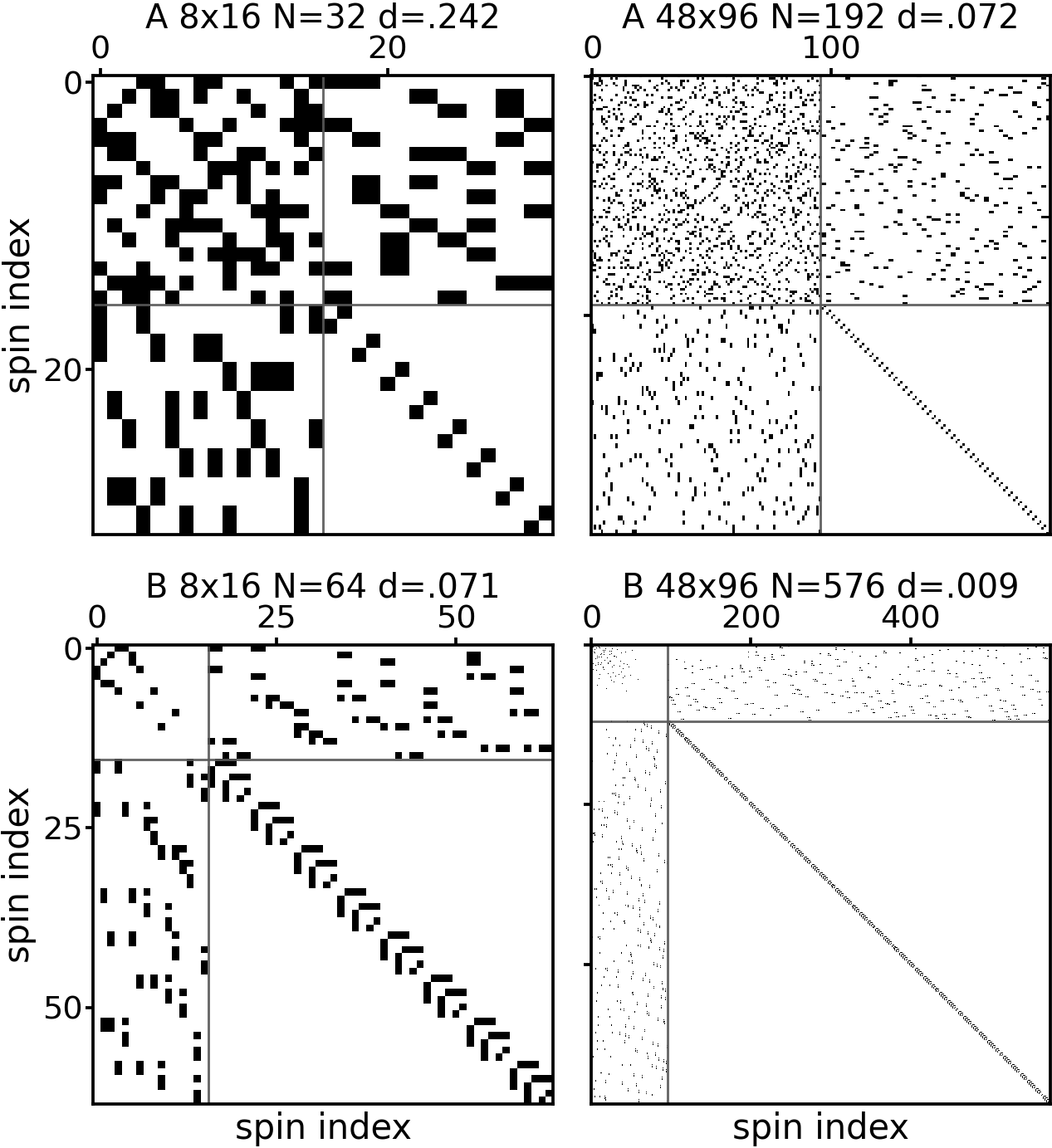}
    \caption{Ising coupling matrices generated by the two parity-check formulations for two code sizes and their corresponding edge densities $d$. Top: Approach A, bottom: Approach B. Left: $\mathbf{H}\in\{0,1\}^{8\times16}$; right: $\mathbf{H}\in\{0,1\}^{48\times96}$. The red lines separate the original code-bit spins from the auxiliary spins.}
    \label{fig:sparsity}
\end{figure}

Fig.~\ref{fig:sparsity} visualize the
coupling graphs induced by the two formulations and quantify the
structural tradeoffs summarised in Table~\ref{tab:comparison}.
Approach~A is the more economical formulation in terms of raw spin
count: each parity check requires only
$B_c=\lceil\log_2(\lfloor d_c/2\rfloor+1)\rceil$ auxiliary spins, giving
a total of $N=n+mB_c$ spins. This makes it attractive whenever the
solver imposes a strict spin budget. The compactness, however, is
obtained by expanding the squared parity penalty, which produces dense
intra-check couplings. Code-bit connectivity reaches $(d_c{-}1+B_c)d_v$. Approach~A can therefore
become difficult to embed/route on hardware with limited fan-out,
even though it uses fewer spins.

Approach~B makes the opposite tradeoff. The chain-product construction
replaces each high-degree parity product with a sequence of local
product constraints, requiring $2(d_c{-}1)$ auxiliary spins per check
and hence $N=n+2m(d_c{-}1)$ total spins. In return, the resulting
coupling graph is substantially sparser. Each code-bit spin acquires
at most $3d_v$ couplings, independent of $d_c$, and every auxiliary
spin interacts with only a constant number of chain neighbors and a
single code-bit spin. This bounded-connectivity structure is well
matched to hardware platforms where local connectivity, wiring, or
embedding overhead is the dominant bottleneck, including quantum
annealers with sparse native connectivity and neuromorphic substrates
with limited cross-core bandwidth.

The price is an enlarged search space and auxiliary-chain degrees of
freedom that must be kept consistent with the code-bit spins, which can affect solver behavior as discussed in Sec.~\ref{sec:solvers}, and an increase in physical area and power since each spin is typically
realized by a dedicated processing element (e.g., a neuron). In
practice, however, contemporary neuromorphic platforms already scale
comfortably beyond the spin counts implied by either formulation. A
single Intel Loihi~2 chip supports up to $\sim\!10^{6}$ neurons and
$1.2\times10^{8}$ synapses across $128$ neuro-cores, with
$\sim\!8\times10^{3}$ neurons available within a single locally-dense
core~\cite{loihi2}, while system-level deployments such as Intel
Hala~Point reach $1.15\times10^{9}$ neurons and $1.28\times10^{11}$
synapses~\cite{halapoint}. For a
IEEE~802.11n $(648,324)$ code, Approach~B requires
$n+2m(d_c{-}1)\approx 4.5\times 10^{3}$ neurons, leaving comfortable
headroom for hundreds of independent decoder instances per chip.

\begin{table}[t]
\centering
\caption{Structural comparison of the two Ising formulations
for a regular $(d_v,d_c)$ code with $n$~code bits and
$M=nd_v/d_c$ checks.}
\label{tab:comparison}
\renewcommand{\arraystretch}{1.15}
\begin{tabular}{lcc}
\hline
 & \textbf{Approach A} & \textbf{Approach B} \\
\hline
Aux.\ spins/check
  & $\lceil\log_2(\lfloor d_c/2\rfloor\!+\!1)\rceil$
  & $2(d_c-1)$ \\
Total spins $N$
  & $n+M B_c$
  & $n+2M(d_c\!-\!1)$ \\
Max.\ connectivity
  & $(d_c\!-\!1+B_c)\,d_v$
  & $3\,d_v$ \\
% Density scaling ($d_c$)
%   & $O(d_c^2)$
%   & $O(d_c)$ \\
% Coupling topology
%   & \multicolumn{2}{c}{fixed offline by $\mathbf{H}$} \\
% Per-block update
%   & \multicolumn{2}{c}{diagonal (LLRs) only} \\
\hline
% \vspace{-20pt}
\end{tabular}
\end{table}

\begin{figure}[h]
    \centering
        \includegraphics[width=\linewidth]{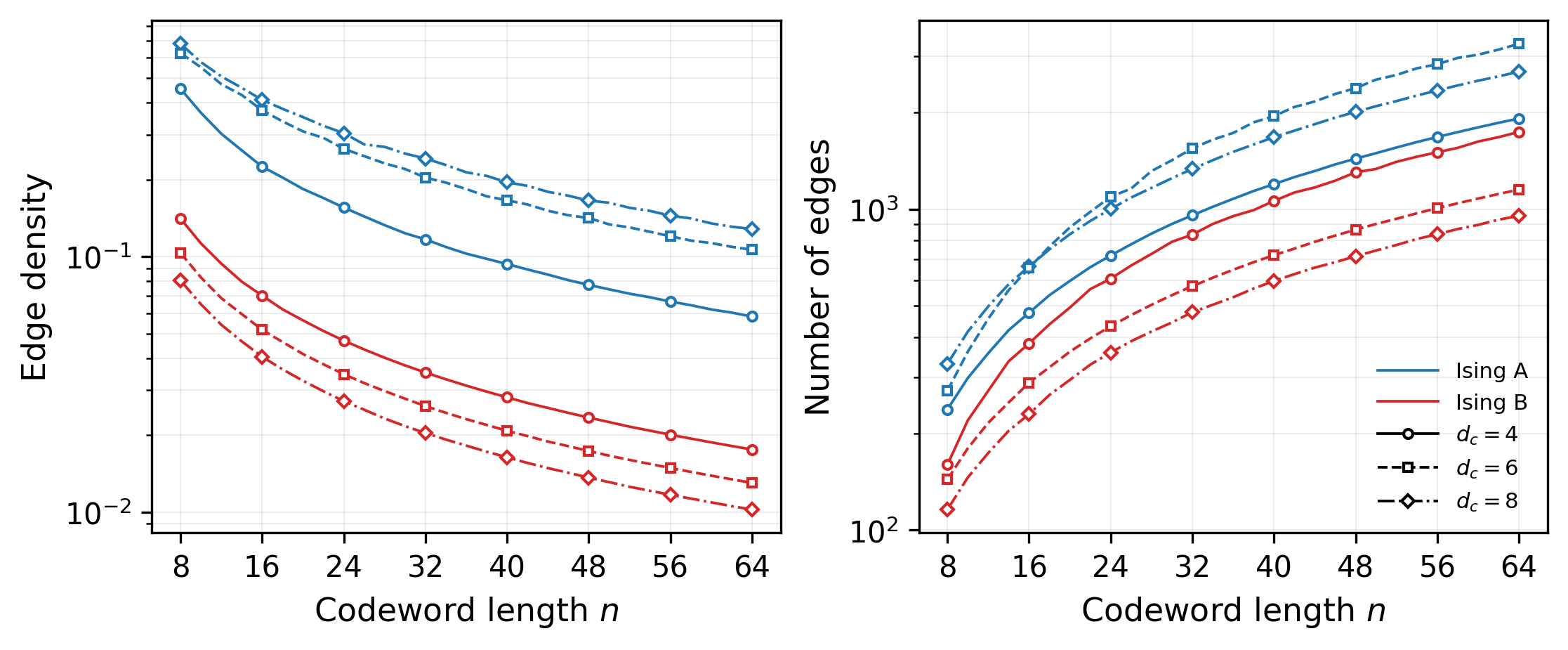}
    \caption{Coupling-graph sparsity of the two formulations as a function of
  codeword length. Left: edge density of the Ising coupling graph. Right: absolute number of
  nonzero couplings}
    \label{fig:density}
\end{figure}

Fig.~\ref{fig:density} evaluates the scaling argument. Right  is the absolute number of non-zero
couplings, which grows with both $n$ and $d_c$ and is consistently
larger for Approach~A. Left, the normalized by the number of
admissible spin pairs. Along the $n$-axis, density decreases for both formulations because the
denominator grows quadratically while couplings induced by a sparse
parity-check matrix grow only linearly. Along the $d_c$-axis, however, the two formulations move in \emph{opposite} directions. Approach~A
adds $\mathcal{O}(d_c^{2})$ couplings on an essentially fixed spin
count, so density rises with $d_c$. Approach~B adds only
$\mathcal{O}(d_c)$ chain couplings but $\mathcal{O}(d_c)$ auxiliary
spins, resulting in a drop of the edge density with~$d_c$.

Together, these observations indicate that the formulation choice tracks the neuromorphic hardware class.
Approach A is more suitable for platforms with locally dense  fabrics, such as crossbar-based designs \cite{jiang_annealing_memristorxbar}, where the dense intra-check couplings or even all-to-all weighted graphs can be
represented within a local array. By contrast, on distributed many-core spiking systems where each coupling contributes to routing and communication cost, Approach B’s bounded-connectivity chain structure is preferable.

% Density figure (placeholder)

\subsection{Programming Overhead}\label{sec:programming}
A practical advantage common to both formulations is that the coupling
topology is fixed by the parity-check matrix $\mathbf{H}$ and by the chosen
penalty weights. The received channel observations enter only through the LLRs,
which modify the local fields in the Ising formulation or equivalently the diagonal terms in the QUBO representation. Therefore, for a fixed code, the off-diagonal couplings can be
calculated offline and reused across received codeblocks, leaving only one scalar
field per code-bit spin to be updated at runtime. 
This is an important distinction from existing Ising formulations of
MIMO detection, where the coupling matrix itself depends on the
channel realisation and must be partially reprogrammed at the channel
coherence rate, incurring runtime preprocessing overhead on a
conventional computing host~\cite{neuromimo}.

\section{Conclusion}\label{sec:conclusion}
This work compared two Ising formulations of ML channel
decoding. Although both formulations can place the ML codeword at the ground
state under sufficient constraint enforcement, our results show that
ground-state correctness alone is not a sufficient criterion. Stronger
penalties improve exactness, but also increase the number of local minima
and shrink the basin of attraction of the ML solution. The two formulations impose different hardware tradeoffs. The squared-penalty formulation minimizes spin count but induces dense intra-check couplings, making it more attractive when dense local fan-in is available. The chain-product formulation increases the number of auxiliary spins but produces a sparse, locally structured graph, making it
better suited to distributed spiking platforms where routing and communication costs dominate. These results show that Ising mappings for channel decoding must be evaluated jointly with the solver dynamics and the intended hardware architecture, providing a basis for the co-design
of neuromorphic FEC accelerators.

\bibliographystyle{IEEEtran}
\bibliography{references.bib}

\end{document}